\documentclass[prc,twocolumn,superscriptaddress,nofootinbib]{revtex4-1}
\usepackage{newtxtext}
\usepackage[varvw,bigdelims]{newtxmath}
\usepackage[colorlinks=true,allcolors=blue]{hyperref}
\usepackage{graphicx}
\usepackage{bm}
\usepackage{amsmath}
\usepackage{xcolor}

\begin{document}
\title{\boldmath 
Line shapes of $\Omega(2012)$ production in the $\Xi \bar K $ and $\Xi \pi \bar K$ decay channels
} 

\author{Natsumi Ikeno}
\email{ikeno@maritime.kobe-u.ac.jp}
\affiliation{Graduate School of Maritime Sciences, Kobe University, Kobe 658-0022, Japan}

\author{Eulogio Oset}
\email{Eulogio.Oset@ific.uv.es}
\affiliation{Departamento de F\'{i}sica Teórica and IFIC, Centro Mixto Universidad de Valencia-CSIC, Institutos de Investigaci\'{o}n de Paterna, Aptdo. 22085, E-46071 Valencia, Spain}

\begin{abstract}
We theoretically study the line shape of the experiment for $\Omega(2012)$ production in the $\Xi^- \pi^+ K^-$ and $\Xi^0 K^-$ decay modes, reflecting the $\Xi^* \bar K$ and $\Xi \bar K$ decay modes, from the perspective that the $\Omega(2012)$ is a molecular state dynamically generated from the interaction of the $\Xi^* \bar K $ and $\Omega \eta$, coupled channels with $\Xi \bar K$ as a decay channel. 
We show the consistency of the picture with the experimental mass distributions, giving support to the molecular picture for the $\Omega(2012)$ state. 
We also call the attention to the sensitivity of the results to the cut imposed on the $\pi \Xi$ invariant mass, and suggest a different method to obtain the ratio $R^{\Xi \pi \bar K}_{\Xi \bar K}$ of the three-body to two-body decay widths used so far as a test of the molecular picture.
We stress that the direct comparison with the experimental mass distributions done here is a more stringent test of the molecular nature of the $\Omega(2012)$ state than the comparison of the ratio $R^{\Xi \pi \bar K}_{\Xi \bar K}$ which has been so far obtained using different criteria.

\end{abstract}

\date{\today}

\maketitle
\section{Introduction}
The Belle Collaboration reported the discovery of a new $\Omega$ state, the $\Omega(2012)$ in Ref.~\cite{Belle:2018mqs} from the decay of $\Upsilon$ states. More recently it has also been observed by the Belle Collaboration in the decay of $\Omega_c$ to $\pi \Omega(2012)$~\cite{Belle:2021gtf}, and by the ALICE collaboration in ultrarelativistic $p p$ collisions in Ref.~\cite{ALICE:2025atb}. A state of these characteristics had been predicted  as a dynamically generated state from the interaction of the octet of pseudoscalar mesons with the baryons of the $\Delta(1232)$ decuplet~\cite{Hofmann:2006qx,Sarkar:2004jh}. In particular, it would correspond to a molecular state of the  $\bar K \Xi^*(1530)$ and $\eta \Omega$. More papers have supported the molecular picture~\cite{Valderrama:2018bmv,Lin:2018nqd,Pavao:2018xub,Huang:2018wth,Lu:2020ste,Ikeno:2020vqv,Ikeno:2022jpe,Liu:2020yen,Zeng:2020och,Lin:2019tex,Liu:2019wdr,Shen:2025xcq,Yu:2026qij}, and normally they also incorporate the $\bar K \Xi$ channel since  the state was observed in this channel. 

 The molecular picture is not unique and there are alternative proposals, like  a $P$-wave excited 3/2$^-$ state of the quark model~\cite{Xiao:2018pwe,Aliev:2018yjo,Aliev:2018syi,Polyakov:2018mow,Arifi:2022ntc,Wang:2022zja,Wang:2018hmi,Zhong:2022cjx,Su:2024lzy}. Some other quark models, however, also claim a molecular nature of the state through the formation of hadron pairs~\cite{Wang:2007bf,Gutsche:2019eoh,Hu:2022pae}. Other works suggest a mixture of $q \bar q$ and molecular components~\cite{Lu:2022puv}. It has also been investigated from the perspective of hamiltonian effective theory in Ref.~\cite{Han:2025gkp}. The Weinberg compositeness condition has also been used in Ref.~\cite{Gutsche:2019eoh} to support the molecular picture. A suggestion to observe it in the  $\psi (3770) \rightarrow \bar{\Omega } \bar{K} \Xi $ and $\psi (3770) \rightarrow \bar{\Omega } \bar{K} \Xi ^*(1530) (\bar{\Omega } \bar{K} \pi \Xi )$ reactions is done in Ref.~\cite{Song:2024ejc}, through $\Omega_b \to J/\psi \Omega^*$ in Ref.~\cite{Wang:2024ozz}, and by looking at correlation functions in Ref.~\cite{Lin:2026ypf}.  Reviews of these ideas can be found in Refs.~\cite{Huang:2023jec,Xie:2024wbd}.

After the discovery, several tests were carried out to support or refute the molecular nature of the $\Omega(2012)$. In this direction, an experiment looking at $\Xi \pi  \bar K$, which would come from the decay of the $\Xi^*(1530) \bar K$ component, was performed in Ref.~\cite{Belle:2019zco}. 
A small fraction with an upper limit of 11.9\%, relative to the $\Xi \bar K$ decay mode, was reported, which was in the limit of disqualifying the molecular pictures~\cite{Lu:2020ste,Ikeno:2020vqv}. 
Yet, a reanalysis of the same experimental data was later performed using different cuts on the $\Xi^- \pi^+$ mass distribution with a new result: $R^{\Xi \pi \bar K}_{\Xi \bar K} = 0.99 \pm 0.26 \pm 0.06$~\cite{Belle:2022mrg}, which is in line with the predictions made in Ref.~\cite{Pavao:2018xub}, stressing the support of these results for the molecular picture.
It is important to stress, however, that different criteria were used in theoretical works and in the two experimental analysis concerning the $R^{\Xi \pi \bar K}_{\Xi \bar K} $. A better comparison can be achieved by directly comparing with the line shapes of the experiment.

  With so much work devoted to this resonance, it looks surprising that none of the works is dealing with the line shape of the $\Omega(2012)$ production in the $\Upsilon$ decays, something considered of extreme importance for a test of consistency of the models~\cite{Tanida_private}. The calculation of the line shapes for $\Xi \bar K$  and $\Xi \pi \bar K$ production in $\Upsilon$ decays, shown in the work of Ref.~\cite{Belle:2022mrg}, is the purpose of the present work.

\section{Formalism}\label{sec:formalism}
\begin{figure}[!hbt]
\begin{center}
\includegraphics[width=0.7\linewidth]{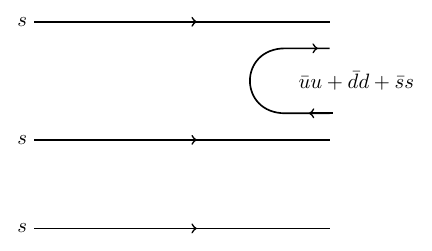}
\caption{
Hadronization of the doorway $sss$ state.}
\label{fig:1}
\end{center}
\end{figure}

We follow the work of Ref.~\cite{Pavao:2018xub}. In order to produce the $\Omega(2012)$ in $\Upsilon$ decays, one assumes that a doorway state of three quarks, $sss$ is formed. The next step is to let it undergo hadronization of an $ss$ pair, as shown in Fig.~\ref{fig:1}.

\begin{figure*}[!bht]
\begin{center}
\includegraphics[width=0.7\linewidth]{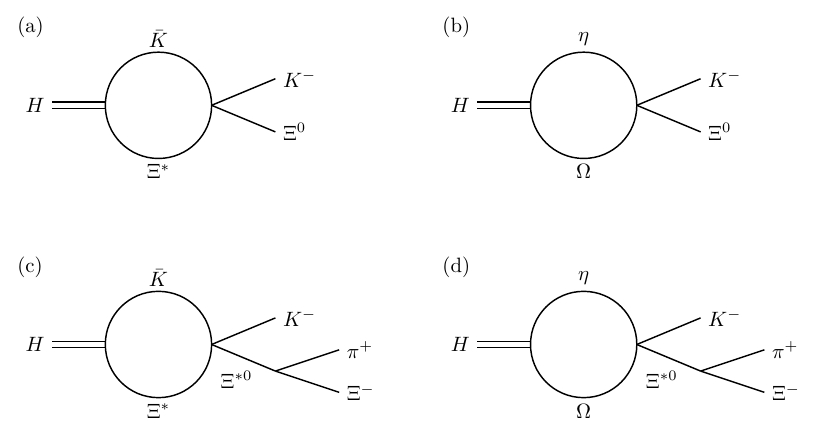}
\caption{
Diagrams involved in the generation of the $\Omega(2012)$ and posterior decay to $K^- \Xi^0$ (diagrams (a) and (b)) and $K^- \pi^+ \Xi^-$ (diagrams (c) and (d)).}
\label{fig:2}
\end{center}
\end{figure*}

The hadronization of Fig.~\ref{fig:1} leads to
\begin{equation}
sss \to H \equiv \sum^{3}_{i=1} s \bar q_i q_i ss,
\end{equation}
identifying $s\bar q_i$ as $\bar K$ or $\eta$, $\eta^\prime$ and $q_i ss$ as $\Xi^* (1530)$ or $\Omega$ states, as done in Ref.~\cite{Pavao:2018xub}, one obtains
\begin{equation}
 H = \frac{1}{\sqrt{3}} K^- \Xi^{*0} +  \frac{1}{\sqrt{3}} K^0 \Xi^{*-} - \frac{1}{\sqrt{3}} \eta \Omega^-,
\end{equation} 
and considering the isospin multiplets $(\bar K^0,\,-K^-)$, $(\Xi^{*0},\Xi^{*-})$,
we have
\begin{equation}
|H\rangle 
=
\sqrt{\frac{2}{3}}\,
|\bar K \, \Xi^{*};\, I=0\rangle
-\frac{1}{\sqrt{3}}
|\Omega \eta\rangle .
\label{eq:H3}
\end{equation}
The production of the $\Omega(2012)$ from the molecular perspective, proceeds through the production of its components, as shown in Eq.~\eqref{eq:H3}, and letting them propagate making a transition to $\bar K \Xi$ and $\bar K \pi \Xi$ in the final states, which are the channels observed in Ref.~\cite{Belle:2022mrg}. This is depicted in Fig.~\ref{fig:2}.

The mechanisms of Fig.~\ref{fig:2} imply that the building blocks of the $\Omega(2012)$ are $\bar K\Xi^{*}$ and $\eta\Omega$, with the $\bar K\Xi$ playing a secondary role, as just a decay channel.
The model of Ref.~\cite{Pavao:2018xub} allows the transition to these channels, which one has in Fig.~\ref{fig:2}~diagrams (a), 2(b). The transition to $K^- \pi^+ \Xi^-$ in this picture proceeds through the transition of the building blocks to $K^- \Xi^{*0}$, followed by the decay of $\Xi^{*0}$ to $\pi^+\Xi^-$.

Analytically, the diagrams of Fig.~\ref{fig:2}(a) and (b) correspond to the transition amplitudes
\begin{equation}
\begin{aligned}
t_{K^- \Xi^0}
=
\mathcal{C}
\Bigg[
&
\sqrt{\frac{2}{3}}\,
G_{\bar K\Xi^{*}}(M_{\rm inv})
\,t_{\bar K\Xi^{*},\,\bar K\Xi}(M_{\rm inv})\,
\frac{1}{\sqrt{2}}
\\
&
-\frac{1}{\sqrt{3}}\,
G_{\eta\Omega}(M_{\rm inv})
\,t_{\eta\Omega,\,\bar K\Xi}(M_{\rm inv})\,
\frac{1}{\sqrt{2}}
\Bigg] ,
\end{aligned}
\label{eq:t2B}
\end{equation}
where $G_i$ are the loop functions for meson-baryon propagation, $t_{ij}$ the transition scattering amplitudes from state $i$ to state $j$, and $M_{\rm inv}$ corresponds to $M_{\rm inv}(K^-\Xi^0)$. $\mathcal{C}$ is a global normalization constant, to be fitted to the number of experimental events.
In Eq.~\eqref{eq:t2B} we have considered the weights of the $\bar K \Xi^{*}$ and $\eta\Omega$ components of $H$ in Eq.~\eqref{eq:H3}, and a Clebsch--Gordan coefficient $\frac{1}{\sqrt{2}}$ for the $K^- \Xi^0$ component of the state $\bar K\Xi$ in $I=0$.

Similarly, the decay of $H$ to $K^- \pi^+ \Xi^-$ of Figs.~\ref{fig:2}~(c) and (d) proceeds through the transition amplitude,
\begin{equation}
\begin{aligned}
t_{K^- \pi^+\Xi^-}
=&\,
\mathcal{C} \Bigg[
\sqrt{\frac{2}{3}}\,
G_{\bar K \Xi^{*}}(\sqrt{s})\,
t_{\bar K \Xi^{*},\,\bar K \Xi^{*}}(\sqrt{s})\,
\frac{1}{\sqrt{2}}
\\
&\quad
-
\frac{1}{\sqrt{3}}\,
G_{\eta\Omega}(\sqrt{s})\,
t_{\eta\Omega,\,\bar K \Xi^{*}}(\sqrt{s})\,
\frac{1}{\sqrt{2}}
\Bigg]
\\
&\cdot
\frac{1}
{M_{\rm inv}(\pi^+\Xi^-)-M_{\Xi^{*0}}
+i\Gamma_{\Xi^{*0}}/2}
\,g_{\Xi^{*0},\,\pi^+\Xi^-} \ \tilde q_\pi\, ,
\end{aligned}
\label{eq:t3B}
\end{equation}
with $\tilde q_\pi$ the $\pi^+$ momentum in the $\Xi^{*0}$ rest frame
\begin{equation}
\tilde q_{\pi^+}
=
\frac{
\lambda^{1/2}
\!\left(
M_{\rm inv}^2(\pi^+\Xi^-),
m_\pi^2,
M_{\Xi^-}^2
\right)
}
{2\,M_{\rm inv}(\pi^+\Xi^-)}.
\label{eq:qpi}
\end{equation}
The coupling $g_{\Xi^{*0},\,\pi^+\Xi^-}$, considering the isospin multiplets
$(\Xi^0,-\Xi^-)$, $(-\pi^+,\pi^0,\pi^-)$, is given by
\begin{equation}
g_{\Xi^{*0},\,\pi^+\Xi^-}
=
\sqrt{\frac{2}{3}}\,
g_{\Xi^{*},\,\pi\Xi},
\end{equation}
where $g_{\Xi^{*},\,\pi\Xi}$ is calculated from the experimental decay width of
$\Xi^* \to \pi \Xi$ as
\begin{equation}
\Gamma_{\Xi^{*},{\rm on}} \equiv \Gamma_{\Xi^{*}\to\pi\Xi}
= \frac{1}{2\pi}
\frac{M_\Xi}{M_{\Xi^{*}}}
g_{\Xi^{*},\,\pi\Xi}^{\,2} \ \tilde q_{\pi,{\rm on}}^{\,3},
\end{equation}
with $\tilde q_{\pi,{\rm on}}$, the value of $\tilde q_\pi$ in Eq.~\eqref{eq:qpi} for $M_{\rm inv}(\pi^+\Xi^-)=M_{\Xi^{*}}$, which provides a coupling
\begin{equation}
g_{\Xi^{*},\,\pi\Xi}
= 0.0044~{\rm MeV}^{-1}.
\end{equation}
In Eq.~\eqref{eq:t3B} the $\Xi^*$ propagator is very close to on shell, and to calculate it we take the width $\Gamma_{\Xi^*}$ energy dependent, as done in Ref.~\cite{Ikeno:2020vqv}, by taking
\begin{equation}
\Gamma_{\Xi^{*}}
\bigl(M_{\rm inv}(\pi^+\Xi^-)\bigr)
=
\Gamma_{\Xi^{*},{\rm on}}
\,
\frac{\tilde q_\pi^{\,3}}
     {\tilde q_{\pi,{\rm on}}^{\,3}}
\,
\Theta\!\left(
M_{\rm inv}(\pi^+\Xi^-)
-m_\pi-M_\Xi
\right).
\end{equation}

The decay width of $H\to K^-\Xi^0$ as a function of the $H$ energy is given by
\begin{equation}
\Gamma_{H \to K^-\Xi^0}(\sqrt{s})
=
\frac{1}{2\pi} \frac{M_\Xi}{\sqrt{s}}\,
|t_{K^-\Xi^0}|^2 \, \tilde q_{K^-},
\end{equation}
with
\begin{equation}
\tilde q_{K^-} = \frac{\lambda^{1/2}(s,m_K^2,M_\Xi^2)}{2\sqrt{s}}.
\end{equation}
The width of $H$ decaying to $K^-\pi^+\Xi^-$ is given by
\begin{equation}
\frac{d\Gamma_{H \to K^-\pi^+\Xi^-}}{dM_{\rm inv}(\pi^+\Xi^-)}=
\frac{1}{(2\pi)^3} \frac{1}{4s}
\,2M_\Xi\,2\sqrt{s}\, |t_{K^-\pi^+\Xi^-}|^2\, p_{K^-}\, \tilde q_\pi ,
\label{eq:dGam_3B}
\end{equation}
with
\begin{equation}
p_{K^-} = \frac{\lambda^{1/2}\!\left(s,m_K^2,M_{\rm inv}^2(\pi^+\Xi^-) \right) }{2\sqrt{s}}.
\end{equation}
To obtain the decay width of $H\to K^-\pi^+\Xi^-$, we integrate Eq.~\eqref{eq:dGam_3B} over $M_{\rm inv}(\pi^+\Xi^-)$ as
\begin{equation}
\Gamma_{H \to K^-\pi^+\Xi^-}(\sqrt{s}) = \int \frac{d\Gamma_{H \to K^-\pi^+\Xi^-}}{dM_{\rm inv}(\pi^+\Xi^-)} \, d M_{\rm inv}(\pi^+\Xi^-). 
\end{equation}
We impose the experimental cut $M_{\rm inv}(\pi^+\Xi^-) < 1517$~MeV, following the analysis of Ref.~\cite{Belle:2022mrg}.

The scattering matrix $t_{ij}$ is calculated as in Refs.~\cite{Pavao:2018xub,Ikeno:2020vqv}, from
\begin{equation}
T=[1-VG]^{-1}V ,
\end{equation}
with $V_{ij}$ the transition potential and $G$ the meson--baryon loop function, with the coupled channels $\bar K\Xi^{*}$, $\eta\Omega$, $\bar K\Xi$, the last one in $d$-wave.
The $G_{\bar K\Xi^{*}}$ function is convolved to account for the width of the $\Xi^{*}$. The value of $q_\text{max}$ used in Ref.~\cite{Pavao:2018xub} is $q_{\rm max}=735$ MeV, by means of which a mass and width of the $\Omega(2012)$ in agreement with experiment were obtained.
In Ref.~\cite{Pavao:2018xub} the transition potential from $\bar K\Xi^{*}$ and $\eta\Omega$ to $\bar K\Xi$ are parametrized by fitting the $\Xi^{*}$ decay width, but the parameters $\alpha$, $\beta$ describing these transitions do not have a unique solution.
We have taken advantage of that fact to estimate uncertainties of our results. 
We, thus, take the pairs $\alpha, \beta$ of Table~5 of Ref.~\cite{Pavao:2018xub},  all of which give a ratio of $\Xi\pi\bar K$ to $\Xi \bar K$ decay widths of close to unity, in agreement with experiment, and evaluate the scattering matrix $t_{ij}$ needed in Eqs.~\eqref{eq:t2B} and~\eqref{eq:t3B}.

\section{Numerical Results}\label{sec:result}
\begin{figure*}[!htb]
\begin{center}
\includegraphics[width=0.95\linewidth]{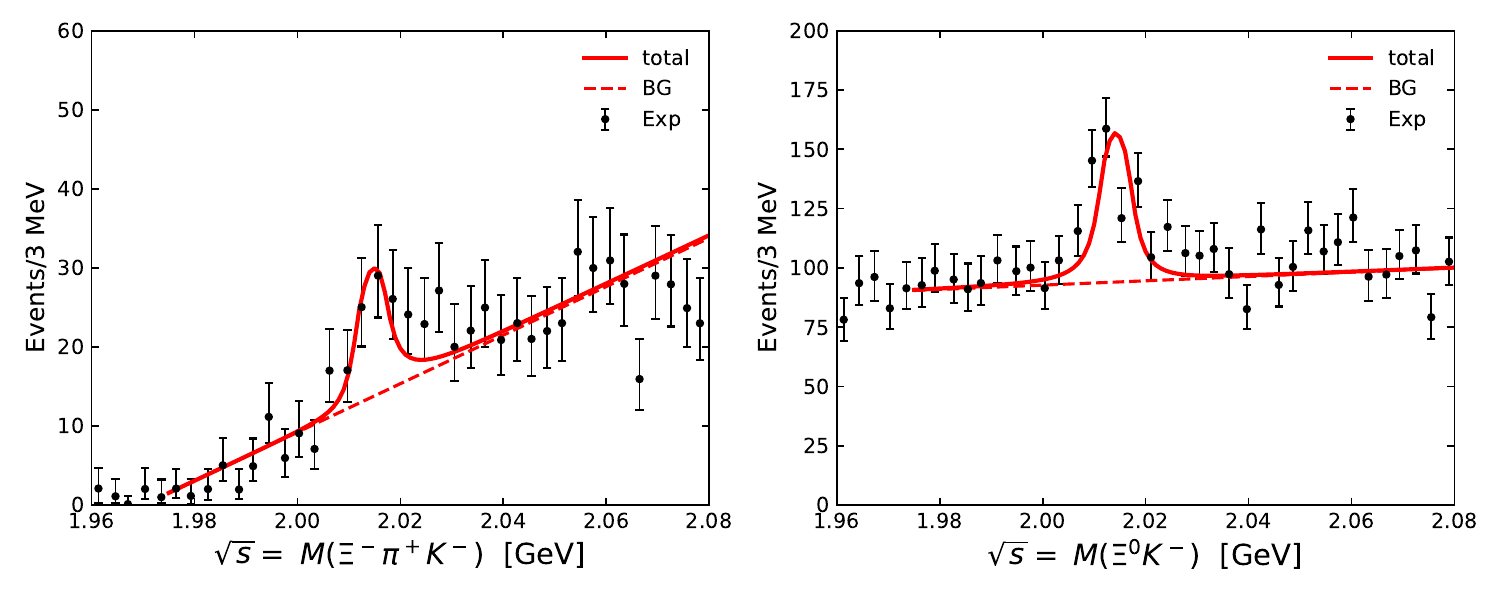}
\caption{
Events in the decay of $H \to K^- \pi^+ \Xi^-$ (Left) and to $K^- \Xi^0$ (Right) as a function of the energy $\sqrt{s}$.
The peaks correspond to the production and posterior decay of the doorway state $H$.
}
\label{fig:3}
\end{center}
\end{figure*}

In Fig.~\ref{fig:3}, 
we show the results for $\Gamma_{H\to K^-\Xi^0}(\sqrt{s})$ and $\Gamma_{H\to K^-\pi^+\Xi^-}(\sqrt{s})$
compared with the experimental results of Ref.~\cite{Belle:2022mrg}.
As in the experimental work, we also take a background coming from events different than the resonant ones. 
We take a background as 
\begin{equation}
N_{\rm BG}(\sqrt{s})
= a_{\rm BG} + b_{\rm BG} ( \sqrt{s}-1.97\,\mathrm{GeV}  ), 
\end{equation}
where the fitted parameters are
$a_{\rm BG}=0$,
$b_{\rm BG}=307~\mathrm{GeV}^{-1}$ for $K^-\pi^+\Xi^-$, and
$a_{\rm BG}=90$,
$b_{\rm BG}=91~\mathrm{GeV}^{-1}$ for $K^-\Xi^0$.

We can see that in both cases the peak distributions are well reproduced using $\alpha = 4.8 \times 10^{-8}$~MeV$^{-3}$, $\beta \simeq 0$.
We have found that there is a correlation between the $\alpha$ and $\beta$ parameters, and different choices give results practically 
equivalent to those obtained with the parameters above.

The merit of the reproduction of the lineshape resides in the fact that Eqs.~\eqref{eq:t2B} and~\eqref{eq:t3B} for the transition amplitudes of $H$ to $K^-\pi^+\Xi^-$ and $K^-\Xi^0$ are well defined in our approach, up to the same global normalization constant.
The use of the $t_{ij}$ and $G_i$ matrices of Ref.~\cite{Pavao:2018xub} guarantees that we have the right mass, widths, and ratio of widths of the $\Omega(2012)$ to $K^-\pi^+\Xi^-$ and $K^-\Xi^0$.
The lineshapes provide additional information, which is tied to the particular weights of the $t_{\bar K\Xi^{*},\,\bar K\Xi}$ and $t_{\eta\Omega,\,\bar K\Xi}$ amplitudes
in Eqs.~\eqref{eq:t2B} and~\eqref{eq:t3B}, and the loop functions $G_{\bar K\Xi^{*}}$ and $G_{\eta\Omega}$ 
which are rather different because the thresholds of the channels are far away
in the transition of the $H$ resonance to the final $K^-\pi^+\Xi^-$ and $K^-\Xi^0$ channels. \\

\section{Discussion} \label{sec:discussion}

It is interesting to make a critical comparison with the analysis and conclusions of the experimental paper~\cite{Belle:2022mrg}. The first thing that we call attention to is the ratio $R^{\Xi \pi K}_{\Xi K}$.  
We can perform the calculation integrating the strength of the two peaks of Fig.~\ref{fig:3}. Note that the scales are different. 
On the other hand, the three body decay shown in the figure (left) is for $\Xi^- \pi^+ K^-$.  
To compare this result with the $\Xi^0 K^-$ of Fig.~\ref{fig:3}(right), we have to reconstruct the $\Xi^{*0} K^-$.  Taking into account the Clebsch Gordan coefficients for the $\Xi^{*0}$ decay into $\Xi^- \pi^+$ and $\Xi^0 \pi^0$, with weights 2/3 and 1/3 respectively, we must multiply the strength of the peak of Fig.~\ref{fig:3}(left) by 3/2 to compare with that of the peak in Fig.~\ref{fig:3}(right). If we integrate from 1975~MeV to 2100~MeV, where the strength of the two resonant peaks has faded, we obtain a ratio $R=0.43$. This is about one half the ratio quoted in Ref.~\cite{Belle:2022mrg}. 
However, if we compare the size of the peaks and width in both cases in Fig.~\ref{fig:3}, the agreement with experiment is good.

\begin{figure}[!htb]
\begin{center}
\includegraphics[width=0.99\linewidth]{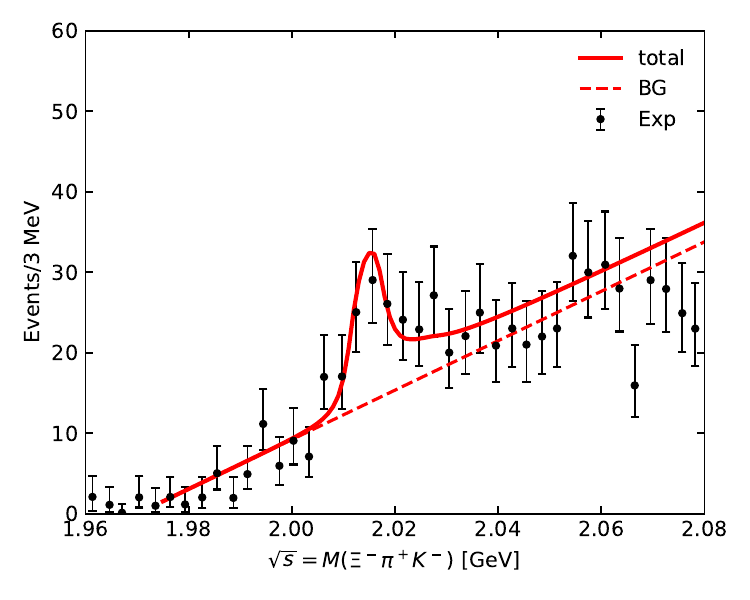}
\caption{
The same as Fig.~\ref{fig:3} (Left) except with the cut $M_{\rm inv}(\pi^+\Xi^-) < 1530$~MeV.
}
\label{fig:4}
\end{center}
\end{figure}

In order to understand the reason of this apparent discrepancy, we must go to the analysis performed in Ref.~\cite{Belle:2022mrg}. First, we call attention to the effect of the cut introduced in Ref.~\cite{Belle:2022mrg}. Indeed, a cut $M_{\rm inv}(\pi^+\Xi^-) < 1517$~MeV was taken there. We make a test to show how sensitive are the results to this cut. For this, we increase the cut to $M_{\rm inv}(\pi^+\Xi^-) < 1530$~MeV and show the new result in Fig.~\ref{fig:4}.
We can see that the results with this new cut are now in better agreement with the data concerning the region to the right of the peak. We also get a hint as to why the errors in that region are relatively large, because of the sensitivity to this cut. 

Next, we see from Ref.~\cite{Belle:2022mrg} that, in order to extract the resonance signal in the three body decay, the authors used a Flatt\'e formula from Ref.~\cite{Hanhart:2010wh}, but this formula does not account for the cut.\footnote{We should mention here that our approach including in a unitary framework all the coupled channels, automatically implements the Flatt\'e behavior, but allows to implement the experimental cuts.} 
As a consequence, they get a shape for the resonance similar to ours in Fig.~\ref{fig:4}, but more centered in the experimental error bars. 
Then, they integrate the fitted function with the Flatt\'e formula of Ref.~\cite{Hanhart:2010wh} on top of their background for a range of invariant masses up to 2200~MeV.
These facts together lead to the ratio 0.99 that is quoted in the experimental paper. 
Therefore, there is not contradiction in the results, it is simply a different way of counting the $\Xi^- \pi^+ K^-$ width. 
The actual direct test of the molecular picture should be done with the comparison with the experimental data, and we can see that the agreement there is good.

\section{Conclusions}\label{sec:conclusions}
We have carried an analysis of the line shape of the experiment for $\Omega(2012)$ production in the $\Xi^- \pi^+ K^-$ and $\Xi^0 K^-$ decay modes, which are related to the $\Xi^* \bar K$ and $\Xi \bar K$ components  of the resonance.  We show that the molecular picture, where the  $\Omega(2012)$ is produced from the interaction of the $\Xi^* \bar K$, $\Omega \eta$, with $\Xi \bar K$ as a decay channel, can reproduce the line shape of the experimental distributions fairly well, both in the shapes and strength of the mass distributions. We showed the sensitivity of the results to a cut in the $M_{\rm inv}(\pi^+\Xi^-)$  done in the experiment, and linked the relatively large experimental errors to the right of the peak in the three body decay to the sensitivity of the results to this cut. We also showed that there was an apparent discrepancy in the ratio $R^{\Xi \pi \bar K}_{\Xi \bar K}$ given by the experimental team and the one we obtained in our analysis of the same experimental data. Yet, the reason of the discrepancies was found in the way of counting 
the $\Xi^- \pi^+ K^-$ in Ref.~\cite{Belle:2022mrg}, integrating the function of in Ref.~\cite{Hanhart:2010wh} that ignored the cut and provided a strength over the chosen background.
In this sense, the present work, providing a direct comparison of the theory with the actual line shape data, is a more stringent test of consistency of the molecular picture with experiment than the comparison of the ratio $R^{\Xi \pi \bar K}_{\Xi \bar K}$ done so far.

\section*{Acknowledgments}
We thank Kiyoshi Tanida for showing the relevance of this work, encouraging us to do it, and providing information on experimental details. 
The work was partly supported by JSPS KAKENHI Grant Number JP24K07020.
This work is partly supported by the Spanish Ministerio de Economia y Competitividad (MINECO) and European FEDER funds under Contracts No. FIS2017-84038-C2-1-P B, PID2020- 112777GB-I00, and by Generalitat Valenciana under contracts PROMETEO/2020/023 and CIPROM/2023/59.This project has received funding from the European Union Horizon 2020 research and innovation programme under the program H2020- INFRAIA2018-1, grant agreement No. 824093 of the STRONG-2020 project. This work is supported by the Spanish Ministerio de Ciencia e Innovacion (MICINN) under contracts PID2020-112777GB-I00, PID2023-147458NBC21 and CEX2023-001292-S.

\bibliography{ref_Omega.bib}

\begin{thebibliography}{41}%
\makeatletter
\providecommand \@ifxundefined [1]{%
 \@ifx{#1\undefined}
}%
\providecommand \@ifnum [1]{%
 \ifnum #1\expandafter \@firstoftwo
 \else \expandafter \@secondoftwo
 \fi
}%
\providecommand \@ifx [1]{%
 \ifx #1\expandafter \@firstoftwo
 \else \expandafter \@secondoftwo
 \fi
}%
\providecommand \natexlab [1]{#1}%
\providecommand \enquote  [1]{``#1''}%
\providecommand \bibnamefont  [1]{#1}%
\providecommand \bibfnamefont [1]{#1}%
\providecommand \citenamefont [1]{#1}%
\providecommand \href@noop [0]{\@secondoftwo}%
\providecommand \href [0]{\begingroup \@sanitize@url \@href}%
\providecommand \@href[1]{\@@startlink{#1}\@@href}%
\providecommand \@@href[1]{\endgroup#1\@@endlink}%
\providecommand \@sanitize@url [0]{\catcode `\\12\catcode `\$12\catcode
  `\&12\catcode `\#12\catcode `\^12\catcode `\_12\catcode `\%12\relax}%
\providecommand \@@startlink[1]{}%
\providecommand \@@endlink[0]{}%
\providecommand \url  [0]{\begingroup\@sanitize@url \@url }%
\providecommand \@url [1]{\endgroup\@href {#1}{\urlprefix }}%
\providecommand \urlprefix  [0]{URL }%
\providecommand \Eprint [0]{\href }%
\providecommand \doibase [0]{http://dx.doi.org/}%
\providecommand \selectlanguage [0]{\@gobble}%
\providecommand \bibinfo  [0]{\@secondoftwo}%
\providecommand \bibfield  [0]{\@secondoftwo}%
\providecommand \translation [1]{[#1]}%
\providecommand \BibitemOpen [0]{}%
\providecommand \bibitemStop [0]{}%
\providecommand \bibitemNoStop [0]{.\EOS\space}%
\providecommand \EOS [0]{\spacefactor3000\relax}%
\providecommand \BibitemShut  [1]{\csname bibitem#1\endcsname}%
\let\auto@bib@innerbib\@empty
\bibitem [{\citenamefont {Yelton}\ \emph {et~al.}(2018)\citenamefont {Yelton}
  \emph {et~al.}}]{Belle:2018mqs}%
  \BibitemOpen
  \bibfield  {author} {\bibinfo {author} {\bibfnamefont {J.}~\bibnamefont
  {Yelton}} \emph {et~al.} (\bibinfo {collaboration} {Belle}),\ }\href
  {\doibase 10.1103/PhysRevLett.121.052003} {\bibfield  {journal} {\bibinfo
  {journal} {Phys. Rev. Lett.}\ }\textbf {\bibinfo {volume} {121}},\ \bibinfo
  {pages} {052003} (\bibinfo {year} {2018})},\ \Eprint
  {http://arxiv.org/abs/1805.09384} {arXiv:1805.09384 [hep-ex]} \BibitemShut
  {NoStop}%
\bibitem [{\citenamefont {Li}\ \emph {et~al.}(2021)\citenamefont {Li} \emph
  {et~al.}}]{Belle:2021gtf}%
  \BibitemOpen
  \bibfield  {author} {\bibinfo {author} {\bibfnamefont {Y.}~\bibnamefont {Li}}
  \emph {et~al.} (\bibinfo {collaboration} {Belle}),\ }\href {\doibase
  10.1103/PhysRevD.104.052005} {\bibfield  {journal} {\bibinfo  {journal}
  {Phys. Rev. D}\ }\textbf {\bibinfo {volume} {104}},\ \bibinfo {pages}
  {052005} (\bibinfo {year} {2021})},\ \Eprint
  {http://arxiv.org/abs/2106.00892} {arXiv:2106.00892 [hep-ex]} \BibitemShut
  {NoStop}%
\bibitem [{\citenamefont {Acharya}\ \emph {et~al.}(2025)\citenamefont {Acharya}
  \emph {et~al.}}]{ALICE:2025atb}%
  \BibitemOpen
  \bibfield  {author} {\bibinfo {author} {\bibfnamefont {S.}~\bibnamefont
  {Acharya}} \emph {et~al.} (\bibinfo {collaboration} {ALICE}),\ }\href
  {\doibase 10.1103/v4mh-3r8z} {\bibfield  {journal} {\bibinfo  {journal}
  {Phys. Rev. D}\ }\textbf {\bibinfo {volume} {112}},\ \bibinfo {pages}
  {092002} (\bibinfo {year} {2025})},\ \Eprint
  {http://arxiv.org/abs/2502.18063} {arXiv:2502.18063 [hep-ex]} \BibitemShut
  {NoStop}%
\bibitem [{\citenamefont {Hofmann}\ and\ \citenamefont
  {Lutz}(2006)}]{Hofmann:2006qx}%
  \BibitemOpen
  \bibfield  {author} {\bibinfo {author} {\bibfnamefont {J.}~\bibnamefont
  {Hofmann}}\ and\ \bibinfo {author} {\bibfnamefont {M.~F.~M.}\ \bibnamefont
  {Lutz}},\ }\href {\doibase 10.1016/j.nuclphysa.2006.07.004} {\bibfield
  {journal} {\bibinfo  {journal} {Nucl. Phys. A}\ }\textbf {\bibinfo {volume}
  {776}},\ \bibinfo {pages} {17} (\bibinfo {year} {2006})},\ \Eprint
  {http://arxiv.org/abs/hep-ph/0601249} {arXiv:hep-ph/0601249} \BibitemShut
  {NoStop}%
\bibitem [{\citenamefont {Sarkar}\ \emph {et~al.}(2005)\citenamefont {Sarkar},
  \citenamefont {Oset},\ and\ \citenamefont {Vicente~Vacas}}]{Sarkar:2004jh}%
  \BibitemOpen
  \bibfield  {author} {\bibinfo {author} {\bibfnamefont {S.}~\bibnamefont
  {Sarkar}}, \bibinfo {author} {\bibfnamefont {E.}~\bibnamefont {Oset}}, \ and\
  \bibinfo {author} {\bibfnamefont {M.~J.}\ \bibnamefont {Vicente~Vacas}},\
  }\href {\doibase 10.1016/j.nuclphysa.2005.01.006} {\bibfield  {journal}
  {\bibinfo  {journal} {Nucl. Phys. A}\ }\textbf {\bibinfo {volume} {750}},\
  \bibinfo {pages} {294} (\bibinfo {year} {2005})},\ \bibinfo {note} {[Erratum:
  Nucl.Phys.A 780, 90--90 (2006)]},\ \Eprint
  {http://arxiv.org/abs/nucl-th/0407025} {arXiv:nucl-th/0407025} \BibitemShut
  {NoStop}%
\bibitem [{\citenamefont {Valderrama}(2018)}]{Valderrama:2018bmv}%
  \BibitemOpen
  \bibfield  {author} {\bibinfo {author} {\bibfnamefont {M.~P.}\ \bibnamefont
  {Valderrama}},\ }\href {\doibase 10.1103/PhysRevD.98.054009} {\bibfield
  {journal} {\bibinfo  {journal} {Phys. Rev. D}\ }\textbf {\bibinfo {volume}
  {98}},\ \bibinfo {pages} {054009} (\bibinfo {year} {2018})},\ \Eprint
  {http://arxiv.org/abs/1807.00718} {arXiv:1807.00718 [hep-ph]} \BibitemShut
  {NoStop}%
\bibitem [{\citenamefont {Lin}\ and\ \citenamefont {Zou}(2018)}]{Lin:2018nqd}%
  \BibitemOpen
  \bibfield  {author} {\bibinfo {author} {\bibfnamefont {Y.-H.}\ \bibnamefont
  {Lin}}\ and\ \bibinfo {author} {\bibfnamefont {B.-S.}\ \bibnamefont {Zou}},\
  }\href {\doibase 10.1103/PhysRevD.98.056013} {\bibfield  {journal} {\bibinfo
  {journal} {Phys. Rev. D}\ }\textbf {\bibinfo {volume} {98}},\ \bibinfo
  {pages} {056013} (\bibinfo {year} {2018})},\ \Eprint
  {http://arxiv.org/abs/1807.00997} {arXiv:1807.00997 [hep-ph]} \BibitemShut
  {NoStop}%
\bibitem [{\citenamefont {Pavao}\ and\ \citenamefont
  {Oset}(2018)}]{Pavao:2018xub}%
  \BibitemOpen
  \bibfield  {author} {\bibinfo {author} {\bibfnamefont {R.}~\bibnamefont
  {Pavao}}\ and\ \bibinfo {author} {\bibfnamefont {E.}~\bibnamefont {Oset}},\
  }\href {\doibase 10.1140/epjc/s10052-018-6329-4} {\bibfield  {journal}
  {\bibinfo  {journal} {Eur. Phys. J. C}\ }\textbf {\bibinfo {volume} {78}},\
  \bibinfo {pages} {857} (\bibinfo {year} {2018})},\ \Eprint
  {http://arxiv.org/abs/1808.01950} {arXiv:1808.01950 [hep-ph]} \BibitemShut
  {NoStop}%
\bibitem [{\citenamefont {Huang}\ \emph {et~al.}(2018)\citenamefont {Huang},
  \citenamefont {Liu}, \citenamefont {Lu}, \citenamefont {Xie},\ and\
  \citenamefont {Geng}}]{Huang:2018wth}%
  \BibitemOpen
  \bibfield  {author} {\bibinfo {author} {\bibfnamefont {Y.}~\bibnamefont
  {Huang}}, \bibinfo {author} {\bibfnamefont {M.-Z.}\ \bibnamefont {Liu}},
  \bibinfo {author} {\bibfnamefont {J.-X.}\ \bibnamefont {Lu}}, \bibinfo
  {author} {\bibfnamefont {J.-J.}\ \bibnamefont {Xie}}, \ and\ \bibinfo
  {author} {\bibfnamefont {L.-S.}\ \bibnamefont {Geng}},\ }\href {\doibase
  10.1103/PhysRevD.98.076012} {\bibfield  {journal} {\bibinfo  {journal} {Phys.
  Rev. D}\ }\textbf {\bibinfo {volume} {98}},\ \bibinfo {pages} {076012}
  (\bibinfo {year} {2018})},\ \Eprint {http://arxiv.org/abs/1807.06485}
  {arXiv:1807.06485 [hep-ph]} \BibitemShut {NoStop}%
\bibitem [{\citenamefont {Lu}\ \emph {et~al.}(2020)\citenamefont {Lu},
  \citenamefont {Zeng}, \citenamefont {Wang}, \citenamefont {Xie},\ and\
  \citenamefont {Geng}}]{Lu:2020ste}%
  \BibitemOpen
  \bibfield  {author} {\bibinfo {author} {\bibfnamefont {J.-X.}\ \bibnamefont
  {Lu}}, \bibinfo {author} {\bibfnamefont {C.-H.}\ \bibnamefont {Zeng}},
  \bibinfo {author} {\bibfnamefont {E.}~\bibnamefont {Wang}}, \bibinfo {author}
  {\bibfnamefont {J.-J.}\ \bibnamefont {Xie}}, \ and\ \bibinfo {author}
  {\bibfnamefont {L.-S.}\ \bibnamefont {Geng}},\ }\href {\doibase
  10.1140/epjc/s10052-020-7944-4} {\bibfield  {journal} {\bibinfo  {journal}
  {Eur. Phys. J. C}\ }\textbf {\bibinfo {volume} {80}},\ \bibinfo {pages} {361}
  (\bibinfo {year} {2020})},\ \Eprint {http://arxiv.org/abs/2003.07588}
  {arXiv:2003.07588 [hep-ph]} \BibitemShut {NoStop}%
\bibitem [{\citenamefont {Ikeno}\ \emph {et~al.}(2020)\citenamefont {Ikeno},
  \citenamefont {Toledo},\ and\ \citenamefont {Oset}}]{Ikeno:2020vqv}%
  \BibitemOpen
  \bibfield  {author} {\bibinfo {author} {\bibfnamefont {N.}~\bibnamefont
  {Ikeno}}, \bibinfo {author} {\bibfnamefont {G.}~\bibnamefont {Toledo}}, \
  and\ \bibinfo {author} {\bibfnamefont {E.}~\bibnamefont {Oset}},\ }\href
  {\doibase 10.1103/PhysRevD.101.094016} {\bibfield  {journal} {\bibinfo
  {journal} {Phys. Rev. D}\ }\textbf {\bibinfo {volume} {101}},\ \bibinfo
  {pages} {094016} (\bibinfo {year} {2020})},\ \Eprint
  {http://arxiv.org/abs/2003.07580} {arXiv:2003.07580 [hep-ph]} \BibitemShut
  {NoStop}%
\bibitem [{\citenamefont {Ikeno}\ \emph {et~al.}(2022)\citenamefont {Ikeno},
  \citenamefont {Liang}, \citenamefont {Toledo},\ and\ \citenamefont
  {Oset}}]{Ikeno:2022jpe}%
  \BibitemOpen
  \bibfield  {author} {\bibinfo {author} {\bibfnamefont {N.}~\bibnamefont
  {Ikeno}}, \bibinfo {author} {\bibfnamefont {W.-H.}\ \bibnamefont {Liang}},
  \bibinfo {author} {\bibfnamefont {G.}~\bibnamefont {Toledo}}, \ and\ \bibinfo
  {author} {\bibfnamefont {E.}~\bibnamefont {Oset}},\ }\href {\doibase
  10.1103/PhysRevD.106.034022} {\bibfield  {journal} {\bibinfo  {journal}
  {Phys. Rev. D}\ }\textbf {\bibinfo {volume} {106}},\ \bibinfo {pages}
  {034022} (\bibinfo {year} {2022})},\ \Eprint
  {http://arxiv.org/abs/2204.13396} {arXiv:2204.13396 [hep-ph]} \BibitemShut
  {NoStop}%
\bibitem [{\citenamefont {Liu}\ \emph {et~al.}(2021)\citenamefont {Liu},
  \citenamefont {Huang}, \citenamefont {Ping},\ and\ \citenamefont
  {Chen}}]{Liu:2020yen}%
  \BibitemOpen
  \bibfield  {author} {\bibinfo {author} {\bibfnamefont {X.}~\bibnamefont
  {Liu}}, \bibinfo {author} {\bibfnamefont {H.}~\bibnamefont {Huang}}, \bibinfo
  {author} {\bibfnamefont {J.}~\bibnamefont {Ping}}, \ and\ \bibinfo {author}
  {\bibfnamefont {D.}~\bibnamefont {Chen}},\ }\href {\doibase
  10.1103/PhysRevC.103.025202} {\bibfield  {journal} {\bibinfo  {journal}
  {Phys. Rev. C}\ }\textbf {\bibinfo {volume} {103}},\ \bibinfo {pages}
  {025202} (\bibinfo {year} {2021})},\ \Eprint
  {http://arxiv.org/abs/2010.15398} {arXiv:2010.15398 [hep-ph]} \BibitemShut
  {NoStop}%
\bibitem [{\citenamefont {Zeng}\ \emph {et~al.}(2020)\citenamefont {Zeng},
  \citenamefont {Lu}, \citenamefont {Wang}, \citenamefont {Xie},\ and\
  \citenamefont {Geng}}]{Zeng:2020och}%
  \BibitemOpen
  \bibfield  {author} {\bibinfo {author} {\bibfnamefont {C.-H.}\ \bibnamefont
  {Zeng}}, \bibinfo {author} {\bibfnamefont {J.-X.}\ \bibnamefont {Lu}},
  \bibinfo {author} {\bibfnamefont {E.}~\bibnamefont {Wang}}, \bibinfo {author}
  {\bibfnamefont {J.-J.}\ \bibnamefont {Xie}}, \ and\ \bibinfo {author}
  {\bibfnamefont {L.-S.}\ \bibnamefont {Geng}},\ }\href {\doibase
  10.1103/PhysRevD.102.076009} {\bibfield  {journal} {\bibinfo  {journal}
  {Phys. Rev. D}\ }\textbf {\bibinfo {volume} {102}},\ \bibinfo {pages}
  {076009} (\bibinfo {year} {2020})},\ \Eprint
  {http://arxiv.org/abs/2006.15547} {arXiv:2006.15547 [hep-ph]} \BibitemShut
  {NoStop}%
\bibitem [{\citenamefont {Lin}\ \emph {et~al.}(2020)\citenamefont {Lin},
  \citenamefont {Lin}, \citenamefont {Wang}, \citenamefont {Zou},\ and\
  \citenamefont {Zou}}]{Lin:2019tex}%
  \BibitemOpen
  \bibfield  {author} {\bibinfo {author} {\bibfnamefont {Y.}~\bibnamefont
  {Lin}}, \bibinfo {author} {\bibfnamefont {Y.-H.}\ \bibnamefont {Lin}},
  \bibinfo {author} {\bibfnamefont {F.}~\bibnamefont {Wang}}, \bibinfo {author}
  {\bibfnamefont {B.}~\bibnamefont {Zou}}, \ and\ \bibinfo {author}
  {\bibfnamefont {B.-S.}\ \bibnamefont {Zou}},\ }\href {\doibase
  10.1103/PhysRevD.102.074025} {\bibfield  {journal} {\bibinfo  {journal}
  {Phys. Rev. D}\ }\textbf {\bibinfo {volume} {102}},\ \bibinfo {pages}
  {074025} (\bibinfo {year} {2020})},\ \Eprint
  {http://arxiv.org/abs/1910.13919} {arXiv:1910.13919 [hep-ph]} \BibitemShut
  {NoStop}%
\bibitem [{\citenamefont {Liu}\ \emph {et~al.}(2020)\citenamefont {Liu},
  \citenamefont {Wang}, \citenamefont {L{\"u}},\ and\ \citenamefont
  {Zhong}}]{Liu:2019wdr}%
  \BibitemOpen
  \bibfield  {author} {\bibinfo {author} {\bibfnamefont {M.-S.}\ \bibnamefont
  {Liu}}, \bibinfo {author} {\bibfnamefont {K.-L.}\ \bibnamefont {Wang}},
  \bibinfo {author} {\bibfnamefont {Q.-F.}\ \bibnamefont {L{\"u}}}, \ and\
  \bibinfo {author} {\bibfnamefont {X.-H.}\ \bibnamefont {Zhong}},\ }\href
  {\doibase 10.1103/PhysRevD.101.016002} {\bibfield  {journal} {\bibinfo
  {journal} {Phys. Rev. D}\ }\textbf {\bibinfo {volume} {101}},\ \bibinfo
  {pages} {016002} (\bibinfo {year} {2020})},\ \Eprint
  {http://arxiv.org/abs/1910.10322} {arXiv:1910.10322 [hep-ph]} \BibitemShut
  {NoStop}%
\bibitem [{\citenamefont {Shen}\ \emph {et~al.}(2026)\citenamefont {Shen},
  \citenamefont {Lu}, \citenamefont {Geng}, \citenamefont {Liu},\ and\
  \citenamefont {Xie}}]{Shen:2025xcq}%
  \BibitemOpen
  \bibfield  {author} {\bibinfo {author} {\bibfnamefont {Q.-H.}\ \bibnamefont
  {Shen}}, \bibinfo {author} {\bibfnamefont {J.-X.}\ \bibnamefont {Lu}},
  \bibinfo {author} {\bibfnamefont {L.-S.}\ \bibnamefont {Geng}}, \bibinfo
  {author} {\bibfnamefont {X.}~\bibnamefont {Liu}}, \ and\ \bibinfo {author}
  {\bibfnamefont {J.-J.}\ \bibnamefont {Xie}},\ }\href {\doibase
  10.1103/y55h-83hg} {\bibfield  {journal} {\bibinfo  {journal} {Phys. Rev. D}\
  }\textbf {\bibinfo {volume} {113}},\ \bibinfo {pages} {036014} (\bibinfo
  {year} {2026})},\ \Eprint {http://arxiv.org/abs/2510.13623} {arXiv:2510.13623
  [hep-ph]} \BibitemShut {NoStop}%
\bibitem [{\citenamefont {Yu}\ \emph {et~al.}(2026)\citenamefont {Yu},
  \citenamefont {Zhang}, \citenamefont {Chen}, \citenamefont {Lian},
  \citenamefont {Wang},\ and\ \citenamefont {Chen}}]{Yu:2026qij}%
  \BibitemOpen
  \bibfield  {author} {\bibinfo {author} {\bibfnamefont {X.}~\bibnamefont
  {Yu}}, \bibinfo {author} {\bibfnamefont {J.-P.}\ \bibnamefont {Zhang}},
  \bibinfo {author} {\bibfnamefont {X.-L.}\ \bibnamefont {Chen}}, \bibinfo
  {author} {\bibfnamefont {D.-K.}\ \bibnamefont {Lian}}, \bibinfo {author}
  {\bibfnamefont {Q.-N.}\ \bibnamefont {Wang}}, \ and\ \bibinfo {author}
  {\bibfnamefont {W.}~\bibnamefont {Chen}},\ }\href {\doibase
  10.1088/1674-1137/ae6da4} {\bibfield  {journal} {\bibinfo  {journal} {Chin.
  Phys. C}\ }\textbf {\bibinfo {volume} {50}},\ \bibinfo {pages} {083106}
  (\bibinfo {year} {2026})},\ \Eprint {http://arxiv.org/abs/2603.03976}
  {arXiv:2603.03976 [hep-ph]} \BibitemShut {NoStop}%
\bibitem [{\citenamefont {Xiao}\ and\ \citenamefont
  {Zhong}(2018)}]{Xiao:2018pwe}%
  \BibitemOpen
  \bibfield  {author} {\bibinfo {author} {\bibfnamefont {L.-Y.}\ \bibnamefont
  {Xiao}}\ and\ \bibinfo {author} {\bibfnamefont {X.-H.}\ \bibnamefont
  {Zhong}},\ }\href {\doibase 10.1103/PhysRevD.98.034004} {\bibfield  {journal}
  {\bibinfo  {journal} {Phys. Rev. D}\ }\textbf {\bibinfo {volume} {98}},\
  \bibinfo {pages} {034004} (\bibinfo {year} {2018})},\ \Eprint
  {http://arxiv.org/abs/1805.11285} {arXiv:1805.11285 [hep-ph]} \BibitemShut
  {NoStop}%
\bibitem [{\citenamefont {Aliev}\ \emph
  {et~al.}(2018{\natexlab{a}})\citenamefont {Aliev}, \citenamefont {Azizi},
  \citenamefont {Sarac},\ and\ \citenamefont {Sundu}}]{Aliev:2018yjo}%
  \BibitemOpen
  \bibfield  {author} {\bibinfo {author} {\bibfnamefont {T.~M.}\ \bibnamefont
  {Aliev}}, \bibinfo {author} {\bibfnamefont {K.}~\bibnamefont {Azizi}},
  \bibinfo {author} {\bibfnamefont {Y.}~\bibnamefont {Sarac}}, \ and\ \bibinfo
  {author} {\bibfnamefont {H.}~\bibnamefont {Sundu}},\ }\href {\doibase
  10.1140/epjc/s10052-018-6375-y} {\bibfield  {journal} {\bibinfo  {journal}
  {Eur. Phys. J. C}\ }\textbf {\bibinfo {volume} {78}},\ \bibinfo {pages} {894}
  (\bibinfo {year} {2018}{\natexlab{a}})},\ \Eprint
  {http://arxiv.org/abs/1807.02145} {arXiv:1807.02145 [hep-ph]} \BibitemShut
  {NoStop}%
\bibitem [{\citenamefont {Aliev}\ \emph
  {et~al.}(2018{\natexlab{b}})\citenamefont {Aliev}, \citenamefont {Azizi},
  \citenamefont {Sarac},\ and\ \citenamefont {Sundu}}]{Aliev:2018syi}%
  \BibitemOpen
  \bibfield  {author} {\bibinfo {author} {\bibfnamefont {T.~M.}\ \bibnamefont
  {Aliev}}, \bibinfo {author} {\bibfnamefont {K.}~\bibnamefont {Azizi}},
  \bibinfo {author} {\bibfnamefont {Y.}~\bibnamefont {Sarac}}, \ and\ \bibinfo
  {author} {\bibfnamefont {H.}~\bibnamefont {Sundu}},\ }\href {\doibase
  10.1103/PhysRevD.98.014031} {\bibfield  {journal} {\bibinfo  {journal} {Phys.
  Rev. D}\ }\textbf {\bibinfo {volume} {98}},\ \bibinfo {pages} {014031}
  (\bibinfo {year} {2018}{\natexlab{b}})},\ \Eprint
  {http://arxiv.org/abs/1806.01626} {arXiv:1806.01626 [hep-ph]} \BibitemShut
  {NoStop}%
\bibitem [{\citenamefont {Polyakov}\ \emph {et~al.}(2019)\citenamefont
  {Polyakov}, \citenamefont {Son}, \citenamefont {Sun},\ and\ \citenamefont
  {Tandogan}}]{Polyakov:2018mow}%
  \BibitemOpen
  \bibfield  {author} {\bibinfo {author} {\bibfnamefont {M.~V.}\ \bibnamefont
  {Polyakov}}, \bibinfo {author} {\bibfnamefont {H.-D.}\ \bibnamefont {Son}},
  \bibinfo {author} {\bibfnamefont {B.-D.}\ \bibnamefont {Sun}}, \ and\
  \bibinfo {author} {\bibfnamefont {A.}~\bibnamefont {Tandogan}},\ }\href
  {\doibase 10.1016/j.physletb.2019.03.054} {\bibfield  {journal} {\bibinfo
  {journal} {Phys. Lett. B}\ }\textbf {\bibinfo {volume} {792}},\ \bibinfo
  {pages} {315} (\bibinfo {year} {2019})},\ \Eprint
  {http://arxiv.org/abs/1806.04427} {arXiv:1806.04427 [hep-ph]} \BibitemShut
  {NoStop}%
\bibitem [{\citenamefont {Arifi}\ \emph {et~al.}(2022)\citenamefont {Arifi},
  \citenamefont {Suenaga}, \citenamefont {Hosaka},\ and\ \citenamefont
  {Oh}}]{Arifi:2022ntc}%
  \BibitemOpen
  \bibfield  {author} {\bibinfo {author} {\bibfnamefont {A.~J.}\ \bibnamefont
  {Arifi}}, \bibinfo {author} {\bibfnamefont {D.}~\bibnamefont {Suenaga}},
  \bibinfo {author} {\bibfnamefont {A.}~\bibnamefont {Hosaka}}, \ and\ \bibinfo
  {author} {\bibfnamefont {Y.}~\bibnamefont {Oh}},\ }\href {\doibase
  10.1103/PhysRevD.105.094006} {\bibfield  {journal} {\bibinfo  {journal}
  {Phys. Rev. D}\ }\textbf {\bibinfo {volume} {105}},\ \bibinfo {pages}
  {094006} (\bibinfo {year} {2022})},\ \Eprint
  {http://arxiv.org/abs/2201.10427} {arXiv:2201.10427 [hep-ph]} \BibitemShut
  {NoStop}%
\bibitem [{\citenamefont {Wang}\ \emph {et~al.}(2023)\citenamefont {Wang},
  \citenamefont {L{\"u}}, \citenamefont {Xie},\ and\ \citenamefont
  {Zhong}}]{Wang:2022zja}%
  \BibitemOpen
  \bibfield  {author} {\bibinfo {author} {\bibfnamefont {K.-L.}\ \bibnamefont
  {Wang}}, \bibinfo {author} {\bibfnamefont {Q.-F.}\ \bibnamefont {L{\"u}}},
  \bibinfo {author} {\bibfnamefont {J.-J.}\ \bibnamefont {Xie}}, \ and\
  \bibinfo {author} {\bibfnamefont {X.-H.}\ \bibnamefont {Zhong}},\ }\href
  {\doibase 10.1103/PhysRevD.107.034015} {\bibfield  {journal} {\bibinfo
  {journal} {Phys. Rev. D}\ }\textbf {\bibinfo {volume} {107}},\ \bibinfo
  {pages} {034015} (\bibinfo {year} {2023})},\ \Eprint
  {http://arxiv.org/abs/2203.04458} {arXiv:2203.04458 [hep-ph]} \BibitemShut
  {NoStop}%
\bibitem [{\citenamefont {Wang}\ \emph {et~al.}(2018)\citenamefont {Wang},
  \citenamefont {Gui}, \citenamefont {L{\"u}}, \citenamefont {Xiao},\ and\
  \citenamefont {Zhong}}]{Wang:2018hmi}%
  \BibitemOpen
  \bibfield  {author} {\bibinfo {author} {\bibfnamefont {Z.-Y.}\ \bibnamefont
  {Wang}}, \bibinfo {author} {\bibfnamefont {L.-C.}\ \bibnamefont {Gui}},
  \bibinfo {author} {\bibfnamefont {Q.-F.}\ \bibnamefont {L{\"u}}}, \bibinfo
  {author} {\bibfnamefont {L.-Y.}\ \bibnamefont {Xiao}}, \ and\ \bibinfo
  {author} {\bibfnamefont {X.-H.}\ \bibnamefont {Zhong}},\ }\href {\doibase
  10.1103/PhysRevD.98.114023} {\bibfield  {journal} {\bibinfo  {journal} {Phys.
  Rev. D}\ }\textbf {\bibinfo {volume} {98}},\ \bibinfo {pages} {114023}
  (\bibinfo {year} {2018})},\ \Eprint {http://arxiv.org/abs/1810.08318}
  {arXiv:1810.08318 [hep-ph]} \BibitemShut {NoStop}%
\bibitem [{\citenamefont {Zhong}\ \emph {et~al.}(2023)\citenamefont {Zhong},
  \citenamefont {Ni}, \citenamefont {Chen}, \citenamefont {Zhong},\ and\
  \citenamefont {Xie}}]{Zhong:2022cjx}%
  \BibitemOpen
  \bibfield  {author} {\bibinfo {author} {\bibfnamefont {H.-H.}\ \bibnamefont
  {Zhong}}, \bibinfo {author} {\bibfnamefont {R.-H.}\ \bibnamefont {Ni}},
  \bibinfo {author} {\bibfnamefont {M.-Y.}\ \bibnamefont {Chen}}, \bibinfo
  {author} {\bibfnamefont {X.-H.}\ \bibnamefont {Zhong}}, \ and\ \bibinfo
  {author} {\bibfnamefont {J.-J.}\ \bibnamefont {Xie}},\ }\href {\doibase
  10.1088/1674-1137/acc9a2} {\bibfield  {journal} {\bibinfo  {journal} {Chin.
  Phys. C}\ }\textbf {\bibinfo {volume} {47}},\ \bibinfo {pages} {063104}
  (\bibinfo {year} {2023})},\ \Eprint {http://arxiv.org/abs/2209.09398}
  {arXiv:2209.09398 [hep-ph]} \BibitemShut {NoStop}%
\bibitem [{\citenamefont {Su}\ \emph {et~al.}(2024)\citenamefont {Su},
  \citenamefont {Chen}, \citenamefont {Gubler},\ and\ \citenamefont
  {Hosaka}}]{Su:2024lzy}%
  \BibitemOpen
  \bibfield  {author} {\bibinfo {author} {\bibfnamefont {N.}~\bibnamefont
  {Su}}, \bibinfo {author} {\bibfnamefont {H.-X.}\ \bibnamefont {Chen}},
  \bibinfo {author} {\bibfnamefont {P.}~\bibnamefont {Gubler}}, \ and\ \bibinfo
  {author} {\bibfnamefont {A.}~\bibnamefont {Hosaka}},\ }\href {\doibase
  10.1103/PhysRevD.110.034007} {\bibfield  {journal} {\bibinfo  {journal}
  {Phys. Rev. D}\ }\textbf {\bibinfo {volume} {110}},\ \bibinfo {pages}
  {034007} (\bibinfo {year} {2024})},\ \Eprint
  {http://arxiv.org/abs/2405.06958} {arXiv:2405.06958 [hep-ph]} \BibitemShut
  {NoStop}%
\bibitem [{\citenamefont {Wang}\ \emph {et~al.}(2007)\citenamefont {Wang},
  \citenamefont {Huang}, \citenamefont {Zhang}, \citenamefont {Yu},\ and\
  \citenamefont {Liu}}]{Wang:2007bf}%
  \BibitemOpen
  \bibfield  {author} {\bibinfo {author} {\bibfnamefont {W.-L.}\ \bibnamefont
  {Wang}}, \bibinfo {author} {\bibfnamefont {F.}~\bibnamefont {Huang}},
  \bibinfo {author} {\bibfnamefont {Z.-Y.}\ \bibnamefont {Zhang}}, \bibinfo
  {author} {\bibfnamefont {Y.-W.}\ \bibnamefont {Yu}}, \ and\ \bibinfo {author}
  {\bibfnamefont {F.}~\bibnamefont {Liu}},\ }\href {\doibase
  10.1088/0253-6102/48/4/025} {\bibfield  {journal} {\bibinfo  {journal}
  {Commun. Theor. Phys.}\ }\textbf {\bibinfo {volume} {48}},\ \bibinfo {pages}
  {695} (\bibinfo {year} {2007})}\BibitemShut {NoStop}%
\bibitem [{\citenamefont {Gutsche}\ and\ \citenamefont
  {Lyubovitskij}(2020)}]{Gutsche:2019eoh}%
  \BibitemOpen
  \bibfield  {author} {\bibinfo {author} {\bibfnamefont {T.}~\bibnamefont
  {Gutsche}}\ and\ \bibinfo {author} {\bibfnamefont {V.~E.}\ \bibnamefont
  {Lyubovitskij}},\ }\href {\doibase 10.1088/1361-6471/abcb9f} {\bibfield
  {journal} {\bibinfo  {journal} {J. Phys. G}\ }\textbf {\bibinfo {volume}
  {48}},\ \bibinfo {pages} {025001} (\bibinfo {year} {2020})},\ \Eprint
  {http://arxiv.org/abs/1912.10894} {arXiv:1912.10894 [hep-ph]} \BibitemShut
  {NoStop}%
\bibitem [{\citenamefont {Hu}\ and\ \citenamefont {Ping}(2022)}]{Hu:2022pae}%
  \BibitemOpen
  \bibfield  {author} {\bibinfo {author} {\bibfnamefont {X.}~\bibnamefont
  {Hu}}\ and\ \bibinfo {author} {\bibfnamefont {J.}~\bibnamefont {Ping}},\
  }\href {\doibase 10.1103/PhysRevD.106.054028} {\bibfield  {journal} {\bibinfo
   {journal} {Phys. Rev. D}\ }\textbf {\bibinfo {volume} {106}},\ \bibinfo
  {pages} {054028} (\bibinfo {year} {2022})},\ \Eprint
  {http://arxiv.org/abs/2207.05598} {arXiv:2207.05598 [hep-ph]} \BibitemShut
  {NoStop}%
\bibitem [{\citenamefont {L{\"u}}\ \emph {et~al.}(2023)\citenamefont {L{\"u}},
  \citenamefont {Nagahiro},\ and\ \citenamefont {Hosaka}}]{Lu:2022puv}%
  \BibitemOpen
  \bibfield  {author} {\bibinfo {author} {\bibfnamefont {Q.-F.}\ \bibnamefont
  {L{\"u}}}, \bibinfo {author} {\bibfnamefont {H.}~\bibnamefont {Nagahiro}}, \
  and\ \bibinfo {author} {\bibfnamefont {A.}~\bibnamefont {Hosaka}},\ }\href
  {\doibase 10.1103/PhysRevD.107.014025} {\bibfield  {journal} {\bibinfo
  {journal} {Phys. Rev. D}\ }\textbf {\bibinfo {volume} {107}},\ \bibinfo
  {pages} {014025} (\bibinfo {year} {2023})},\ \Eprint
  {http://arxiv.org/abs/2212.02783} {arXiv:2212.02783 [hep-ph]} \BibitemShut
  {NoStop}%
\bibitem [{\citenamefont {Han}\ \emph {et~al.}(2025)\citenamefont {Han},
  \citenamefont {Liu}, \citenamefont {Leinweber},\ and\ \citenamefont
  {Thomas}}]{Han:2025gkp}%
  \BibitemOpen
  \bibfield  {author} {\bibinfo {author} {\bibfnamefont {F.-C.}\ \bibnamefont
  {Han}}, \bibinfo {author} {\bibfnamefont {Z.-W.}\ \bibnamefont {Liu}},
  \bibinfo {author} {\bibfnamefont {D.~B.}\ \bibnamefont {Leinweber}}, \ and\
  \bibinfo {author} {\bibfnamefont {A.~W.}\ \bibnamefont {Thomas}},\ }\href
  {\doibase 10.1103/xdbd-v5hb} {\bibfield  {journal} {\bibinfo  {journal}
  {Phys. Rev. D}\ }\textbf {\bibinfo {volume} {112}},\ \bibinfo {pages}
  {L051503} (\bibinfo {year} {2025})},\ \Eprint
  {http://arxiv.org/abs/2507.06682} {arXiv:2507.06682 [hep-ph]} \BibitemShut
  {NoStop}%
\bibitem [{\citenamefont {Song}\ \emph {et~al.}(2024)\citenamefont {Song},
  \citenamefont {Liang}, \citenamefont {Xiao}, \citenamefont {Dias},\ and\
  \citenamefont {Oset}}]{Song:2024ejc}%
  \BibitemOpen
  \bibfield  {author} {\bibinfo {author} {\bibfnamefont {J.}~\bibnamefont
  {Song}}, \bibinfo {author} {\bibfnamefont {W.-H.}\ \bibnamefont {Liang}},
  \bibinfo {author} {\bibfnamefont {C.-W.}\ \bibnamefont {Xiao}}, \bibinfo
  {author} {\bibfnamefont {J.~M.}\ \bibnamefont {Dias}}, \ and\ \bibinfo
  {author} {\bibfnamefont {E.}~\bibnamefont {Oset}},\ }\href {\doibase
  10.1140/epjc/s10052-024-13710-9} {\bibfield  {journal} {\bibinfo  {journal}
  {Eur. Phys. J. C}\ }\textbf {\bibinfo {volume} {84}},\ \bibinfo {pages}
  {1311} (\bibinfo {year} {2024})},\ \Eprint {http://arxiv.org/abs/2410.23204}
  {arXiv:2410.23204 [hep-ph]} \BibitemShut {NoStop}%
\bibitem [{\citenamefont {Wang}\ \emph {et~al.}(2025)\citenamefont {Wang},
  \citenamefont {Wang}, \citenamefont {Hsiao},\ and\ \citenamefont
  {Zhong}}]{Wang:2024ozz}%
  \BibitemOpen
  \bibfield  {author} {\bibinfo {author} {\bibfnamefont {K.-L.}\ \bibnamefont
  {Wang}}, \bibinfo {author} {\bibfnamefont {J.}~\bibnamefont {Wang}}, \bibinfo
  {author} {\bibfnamefont {Y.-K.}\ \bibnamefont {Hsiao}}, \ and\ \bibinfo
  {author} {\bibfnamefont {X.-H.}\ \bibnamefont {Zhong}},\ }\href {\doibase
  10.1103/4g99-psnm} {\bibfield  {journal} {\bibinfo  {journal} {Phys. Rev. D}\
  }\textbf {\bibinfo {volume} {111}},\ \bibinfo {pages} {114028} (\bibinfo
  {year} {2025})},\ \Eprint {http://arxiv.org/abs/2412.02464} {arXiv:2412.02464
  [hep-ph]} \BibitemShut {NoStop}%
\bibitem [{\citenamefont {Lin}\ \emph {et~al.}(2026)\citenamefont {Lin},
  \citenamefont {Encarnaci{\'o}n}, \citenamefont {Feijoo},\ and\ \citenamefont
  {Albaladejo}}]{Lin:2026ypf}%
  \BibitemOpen
  \bibfield  {author} {\bibinfo {author} {\bibfnamefont {J.-X.}\ \bibnamefont
  {Lin}}, \bibinfo {author} {\bibfnamefont {P.}~\bibnamefont
  {Encarnaci{\'o}n}}, \bibinfo {author} {\bibfnamefont {A.}~\bibnamefont
  {Feijoo}}, \ and\ \bibinfo {author} {\bibfnamefont {M.}~\bibnamefont
  {Albaladejo}},\ }\href@noop {} {\  (\bibinfo {year} {2026})},\ \Eprint
  {http://arxiv.org/abs/2603.18610} {arXiv:2603.18610 [hep-ph]} \BibitemShut
  {NoStop}%
\bibitem [{\citenamefont {Huang}\ \emph {et~al.}(2023)\citenamefont {Huang},
  \citenamefont {Deng}, \citenamefont {Liu}, \citenamefont {Tan},\ and\
  \citenamefont {Ping}}]{Huang:2023jec}%
  \BibitemOpen
  \bibfield  {author} {\bibinfo {author} {\bibfnamefont {H.}~\bibnamefont
  {Huang}}, \bibinfo {author} {\bibfnamefont {C.}~\bibnamefont {Deng}},
  \bibinfo {author} {\bibfnamefont {X.}~\bibnamefont {Liu}}, \bibinfo {author}
  {\bibfnamefont {Y.}~\bibnamefont {Tan}}, \ and\ \bibinfo {author}
  {\bibfnamefont {J.}~\bibnamefont {Ping}},\ }\href {\doibase
  10.3390/sym15071298} {\bibfield  {journal} {\bibinfo  {journal} {Symmetry}\
  }\textbf {\bibinfo {volume} {15}},\ \bibinfo {pages} {1298} (\bibinfo {year}
  {2023})}\BibitemShut {NoStop}%
\bibitem [{\citenamefont {Xie}\ and\ \citenamefont {Geng}(2024)}]{Xie:2024wbd}%
  \BibitemOpen
  \bibfield  {author} {\bibinfo {author} {\bibfnamefont {J.-J.}\ \bibnamefont
  {Xie}}\ and\ \bibinfo {author} {\bibfnamefont {L.-S.}\ \bibnamefont {Geng}},\
  }\href {\doibase 10.1088/0256-307X/41/8/081402} {\bibfield  {journal}
  {\bibinfo  {journal} {Chin. Phys. Lett.}\ }\textbf {\bibinfo {volume} {41}},\
  \bibinfo {pages} {081402} (\bibinfo {year} {2024})},\ \Eprint
  {http://arxiv.org/abs/2406.17481} {arXiv:2406.17481 [hep-ph]} \BibitemShut
  {NoStop}%
\bibitem [{\citenamefont {Jia}\ \emph {et~al.}(2019)\citenamefont {Jia} \emph
  {et~al.}}]{Belle:2019zco}%
  \BibitemOpen
  \bibfield  {author} {\bibinfo {author} {\bibfnamefont {S.}~\bibnamefont
  {Jia}} \emph {et~al.} (\bibinfo {collaboration} {Belle}),\ }\href {\doibase
  10.1103/PhysRevD.100.032006} {\bibfield  {journal} {\bibinfo  {journal}
  {Phys. Rev. D}\ }\textbf {\bibinfo {volume} {100}},\ \bibinfo {pages}
  {032006} (\bibinfo {year} {2019})},\ \Eprint
  {http://arxiv.org/abs/1906.00194} {arXiv:1906.00194 [hep-ex]} \BibitemShut
  {NoStop}%
\bibitem [{\citenamefont {Jia}\ \emph {et~al.}(2025)\citenamefont {Jia} \emph
  {et~al.}}]{Belle:2022mrg}%
  \BibitemOpen
  \bibfield  {author} {\bibinfo {author} {\bibfnamefont {S.}~\bibnamefont
  {Jia}} \emph {et~al.} (\bibinfo {collaboration} {Belle}),\ }\href {\doibase
  10.1016/j.physletb.2024.139224} {\bibfield  {journal} {\bibinfo  {journal}
  {Phys. Lett. B}\ }\textbf {\bibinfo {volume} {860}},\ \bibinfo {pages}
  {139224} (\bibinfo {year} {2025})},\ \Eprint
  {http://arxiv.org/abs/2207.03090} {arXiv:2207.03090 [hep-ex]} \BibitemShut
  {NoStop}%
\bibitem [{\citenamefont {Tanida}()}]{Tanida_private}%
  \BibitemOpen
  \bibfield  {author} {\bibinfo {author} {\bibfnamefont {K.}~\bibnamefont
  {Tanida}},\ }\href@noop {} {}\bibinfo {note} {Private
  communication}\BibitemShut {NoStop}%
\bibitem [{\citenamefont {Hanhart}\ \emph {et~al.}(2010)\citenamefont
  {Hanhart}, \citenamefont {Kalashnikova},\ and\ \citenamefont
  {Nefediev}}]{Hanhart:2010wh}%
  \BibitemOpen
  \bibfield  {author} {\bibinfo {author} {\bibfnamefont {C.}~\bibnamefont
  {Hanhart}}, \bibinfo {author} {\bibfnamefont {Y.~S.}\ \bibnamefont
  {Kalashnikova}}, \ and\ \bibinfo {author} {\bibfnamefont {A.~V.}\
  \bibnamefont {Nefediev}},\ }\href {\doibase 10.1103/PhysRevD.81.094028}
  {\bibfield  {journal} {\bibinfo  {journal} {Phys. Rev. D}\ }\textbf {\bibinfo
  {volume} {81}},\ \bibinfo {pages} {094028} (\bibinfo {year} {2010})},\
  \Eprint {http://arxiv.org/abs/1002.4097} {arXiv:1002.4097 [hep-ph]}
  \BibitemShut {NoStop}%
\end{thebibliography}%

\end{document}